\documentclass[aps,prl,showpacs,floats, twocolumn,floats,superscriptaddress,floatfix]{revtex4}
\usepackage{bm}
\usepackage{times}
\usepackage{verbatim}
\usepackage{graphicx}
\usepackage{graphics,epsfig}
\usepackage{theorem}
\usepackage{makeidx}
\usepackage{amsmath}
\usepackage{epic}
\usepackage{amscd}
\usepackage{bbm}
\usepackage{xy}
\usepackage{amssymb}
\usepackage{epsfig}

\usepackage{amsbsy}

\def\bs{\boldsymbol}
\def\gsim{\;\rlap{\lower 2.5pt
\hbox{$\sim$}}\raise 1.5pt\hbox{$>$}\;}
\def\lsim{\;\rlap{\lower 2.5pt
\hbox{$\sim$}}\raise 1.5pt\hbox{$<$}\;}

\begin{document}

\newif\iffigs 
\figstrue
\iffigs \fi
\def\drawing #1 #2 #3 {
\begin{center}
\setlength{\unitlength}{1mm}
\begin{picture}(#1,#2)(0,0)
\put(0,0){\framebox(#1,#2){#3}}
\end{picture}
\end{center} }

\title{Passive Scalar Structures in Supersonic Turbulence} 
\author{Liubin Pan}
\author{Evan Scannapieco}
\affiliation{School of Earth and Space Exploration,  Arizona
  State University, P.O.  Box 871404, Tempe, AZ, 85287-1404.}

\begin{abstract}

We conduct a systematic numerical study 
of passive scalar structures in supersonic turbulent 
flows. We find that the degree of intermittency 
in the scalar structures increases only slightly 
as the flow changes from transonic to highly 
supersonic, while the velocity structures 
become significantly more intermittent. 
This difference is due to the absence 
of shock-like discontinuities in the scalar field. 
The structure functions of the scalar field 
are well described by the intermittency 
model of She and L\'{e}v\^{e}que 
[Phys.\ Rev.\ Lett.\ {\bf 72}, 336 (1994)], and the 
most intense scalar structures are found 
to be sheet-like at all Mach numbers.

\end{abstract}

\maketitle

Supersonic turbulent motions have been 
observed over a wide range of length 
scales in the interstellar medium~\cite{lar81}, 
and the mixing of heavy elements 
released from stellar winds and supernovae 
occurs in such a supersonic turbulent 
environment. Thus understanding passive 
scalar physics in supersonic turbulence is  
crucial for the interpretation of many observational results 
concerning cosmic chemical abundances.
The scalar field we study here is the concentration 
field, $\theta(\bs{x},t)$, of passive tracers.

The statistical approach commonly used 
to study structures in turbulent systems is to 
analyze structure functions (SFs), defined 
as $S^{\rm v}_p (r) \equiv \langle |\delta v (r)|^p \rangle$ 
for the velocity field, or 
$S^{\theta}_p(r) \equiv \langle |\delta \theta (r)|^p \rangle$ 
for the scalar field. 
Here $\delta v (r)$ and $\delta \theta (r)$ are, 
respectively, the longitudinal velocity
increment and the scalar increment
over a distance $r$. 
In the scale range
where the non-linear interactions prevail, 
the SFs are expected to follow power laws, 
$S^{\rm v}_p (r) \propto   r^{\xi_{\rm v} (p)}$,
and  $S^{\theta}_p(r)  \propto  r^{\xi_{\theta}(p)}$.  

The velocity scaling exponents $\xi_{\rm v}(p)$ 
in incompressible turbulence 
depart significantly from the prediction, $\xi_{\rm v} (p) = p/3$, 
by the similarity hypothesis of Kolmogorov at high 
orders ($p >3$), a phenomenon known as anomalous 
scaling or intermittency~\cite{fri95}. 
The physical origin of the departure is the strong 
spatial fluctuations of the energy dissipation rate,  
which provides intermittency corrections to $\xi_{\rm v}(p)$. 
Perhaps the most successful intermittency model is 
the one by She and  L\'{e}v\^{e}que (hereafter 
the SL model)~\cite{she94}, and its prediction for $\xi_{\rm v} (p)$ in 
incompressible turbulence agrees with 
experimental results at high accuracy~\cite{she95}.

In supersonic turbulence, the existence of 
velocity shocks gives more intermittent velocity 
scalings, and  
the degree of intermittency increases with the 
Mach number, $M$. Padoan {\it et al}.~\cite{pad04} 
showed that the velocity scaling in supersonic flows 
can be unified using the SL model with a parameter, 
the dimension of the most intense velocity 
structures (MIVSs), that varies with $M$. These structures 
make a transition from filamentary (vortex tubes) 
at small $M$ to sheet-like (shocks) at large $M$.   

Studies of passive scalars in incompressible 
turbulence found that scalar SFs also have 
anomalous scaling behaviors, 
and that scalar structures are more intermittent than the velocity 
structures~\cite{wat04}.  
The most intense scalar structures (MISSs) 
are sheet-like, corresponding to the observed cliffs 
in the scalar field~\cite{pum94, rui96, lev99,shr00}. 
In this Rapid Communication, we conduct a systematic 
numerical study of passive scalar structures 
in supersonic turbulence.

The classic theory for turbulent mixing~\cite{obu49} assumes,
\begin{equation}
\delta \theta(r)^2 \simeq \bar {\epsilon}_\theta  \frac{r} {\delta v(r)}
\label{eq:obukhov}
\end{equation}     
where $\bar {\epsilon}_\theta$ is the average scalar 
dissipation rate. Refining eq.\ (\ref{eq:obukhov}) to 
account for the fluctuations in the scalar dissipation rate 
gives $\delta \theta(r)^2 \simeq  \epsilon_\theta(r)  \frac{r} {\delta v(r)}$ 
where $\epsilon_{\theta}(r)$ is the dissipation 
rate averaged over a scale $r$. 
The quantity, $\delta v(r) \delta \theta(r)^2$, associated with the scalar 
cascade is of special interest~\cite{lev99}. Defining  ``mixed" SFs as 
$S^{\rm m}_p (r) \equiv \langle |\delta v(r)  \delta \theta(r)^2|^{p/3} \rangle$, 
the intermittency corrections to the mixed structures 
come completely from the fluctuations in the scalar 
dissipation rate. 
We study the velocity,  the mixed, and the scalar SFs, 
denoted as  $S^{\rm v}_p(r)$,  $S^{\rm m}_p(r)$ and $S^{\theta}_p(r)$, 
respectively. We use super- or subscripts, 
${\rm v}$, ${\rm m}$ and $\theta$ to specify each case,  
and when no such super- or sub- scripts are used, 
the discussion is general, referring to all the three cases.    

We use the 512$^3$ simulation data from 
Pan and Scannapieco~\cite{pan10}, who 
simulated isothermal hydrodynamic turbulent 
flows at 6 Mach numbers ($M=0.9$, 1.4, 2.1, 3.0, 4.6, and 6.1) 
using the FLASH code (version 3.2)~\cite{fry00}. 
We integrated the advection equation for the concentration field, 
\begin{equation}
\partial_t \theta + \bs{v} \cdot \nabla \theta = S(\bs{x}, t), 
\end{equation}  
where $S(\bs{x}, t)$ 
represents the tracer sources. In each flow 
we evolved three independent scalars with statistically 
equivalent source terms. Averaging over the three scalars gives more 
accurate measurements. Theoretical studies 
predicted the existence of an inverse scalar cascade 
in highly compressible turbulence~\cite{fal01}, which, however, was 
not observed in our simulations, 
as the critical compressibility for the inverse cascade 
is not reached even at $M=6.1$. Several lines 
of evidence for a direct scalar cascade at all $M$ 
are given in Ref.~\cite{pan10}.  

The viscous and diffusion terms are not 
explicitly included in our simulations, and 
the dissipation of kinetic energy and 
the scalar variance is through numerical diffusion. 
With the dissipation rates determined 
from their balance with the forcing rates, 
we estimated the effective viscosity and 
diffusivity using the relations between the 
dissipation rates and the velocity/scalar gradients. The Taylor 
Reynolds number is estimated to be $\simeq 250$ in the $M=0.9$ flow. 
It decreases to $\simeq 140$ at $M=6.1$ because the code induces
a larger effective viscosity to stabilize stronger shocks. 
One important issue in passive scalar physics is the 
effect of the Schmidt number, $Sc$ (see Ref.\ \cite{ant03} for a 
detailed review). Our estimated viscosity and diffusivity 
give an effective $Sc \simeq 1$ at all $M$. 
However, we point out that uncertainty exists 
in the estimate, and the $Sc$ effect in our 
simulations cannot be exactly evaluated. 
The $Sc$ effect for mixing in supersonic 
turbulence needs to be studied in future 
simulations including viscous and diffusion 
terms. 

We first consider the 3rd order SFs, which 
will be taken as references in our scaling 
analysis below. 
The measured values for $\xi_{\rm v}(3)$, $\xi_{\rm m}(3)$ 
and $\xi_{\theta}(3)$  
are listed in Table 1. 
At $M=0.9$, we find that $\xi_{\rm v}(3)$ 
and $\xi_{\rm m}(3)$ are close to 
unity, consistent with Kolmogorov's 4/5 law~\cite{fri95}, 
and the 4/3 law by Yaglom ~\cite{yag49} in 
incompressible turbulence.   
As $M$ increases to $2$, $\xi_{\rm v} (3)$ 
for the velocity field increases, and then 
saturates at 1.22 for $M\gsim 3$,  close to 
that found in Ref.\ \cite{kri07}. 

Eq.\ (\ref{eq:obukhov}) suggests that 
$\xi_{\rm m}(3)$ is equal to 1 at all Mach 
numbers. In fact, we find that $\xi_{\rm m} (3)$ 
is near 1 at each $M$, supporting the general validity of 
the classic theory for passive scalar 
physics in supersonic turbulence. 
However, we note that $\xi_{\rm m}$ 
starts to increase continuously at $M \gsim 3$, 
meaning that eq.\ (\ref{eq:obukhov}) is not 
exactly obeyed at large $M$. This is likely 
to be caused by the effect of compressible 
modes on scalar structures, which is not reflected 
in eq.\ (\ref{eq:obukhov}). 

For scalar structures, we find $\xi_{\theta}(3)=0.87$ at $M=0.9$, 
which is the same as that obtained in Ref.~\cite{wat04} 
for the incompressible case. The exponent $\xi_{\theta}(3)$ 
first decreases with increasing $M$, and then starts to increase 
at $M \gsim 3$, similar to the behavior of $\xi_{\theta}(2)$ 
found in Ref.~\cite{pan10}. A detailed explanation for this 
behavior is given in Ref.~\cite{pan10}.

Measuring scaling exponents $\xi(p)$ 
at high orders is difficult due to the 
limited inertial range in our simulations. 
Here we exploit the concept of 
extended self-similarity~\cite{ben93}. 
Plotting the SFs at all orders against the 3rd order 
ones gives more extended power-law ranges, 
allowing more accurate measurements. 
We thus measure the scaling exponents, 
$\zeta (p)$, relative to the 3rd order SFs, 
defined as $S_{p}(r) \propto \left( S_{3} (r) \right)^{\zeta (p)}$. 
By definition, $\zeta (p) = \xi (p)/ \xi (3)$ and $\zeta (3) =1$.  

Fig.\ 1 shows the measurement of $\zeta(p)$ for the scalar SFs 
at $M = 6$. The data points in the figure, from left to right, 
correspond to $r=4$ to $196$ (in units of the size of 
the computation cells). An extended power-law range is seen at low to 
intermediate orders, 
and the range extends into the dissipative scales~\cite{ben93}.
At large $p$, the power-law range becomes 
smaller, and we calculated the slope using 
least-square fits with the central 6 points, 
corresponding to $r \in [12, 64]$.      
An examination of the probability distribution of 
$\delta v (r)$ and $\delta \theta(r)$ shows that 
our data have good statistics at orders 
$p$ up to 10 for $M \lsim 3$. At $M = 4.5$, 
and $6.1$, the statistics for $p \simeq 10$ 
appears to be sufficient, but the measurement uncertainty becomes 
larger, especially for the velocity SFs. 
One reason for this is that, at larger $M$, the 
effective numerical viscosity in our simulations is larger, 
giving a narrower inertial range.  

In Fig.\ 2, we plot the measured exponents, 
$\zeta (p)$, for the three kinds of SFs 
at $M=0.9$ and $M=6.1$. 
In the $M=0.9$ flow, $\zeta (p)$ for all the 
three SFs agrees well  with the results 
for incompressible turbulence (within 4\% at all orders, $p$)
~\cite{por02, ben93, lev99, wat04}. 
At $M=0.9$, $\zeta_{\theta} (p)$ is smaller 
than both $\zeta_{\rm m}(p)$ and $\zeta_{\rm v}(p)$, 
indicating the scalar structures are the 
most intermittent in the incompressible limit.  
As $M$ increases to $6.1$,  the relative degree of 
intermittency is reversed, with the velocity field becoming 
more intermittent than the scalar field. 
The transition between these two regimes occurs at $M=2$, 
where the $\zeta_{\rm v}(p)$ and $\zeta_{\theta}(p)$ 
curves are very close to each other.

\begin{figure}
\includegraphics[width=0.5\textwidth]{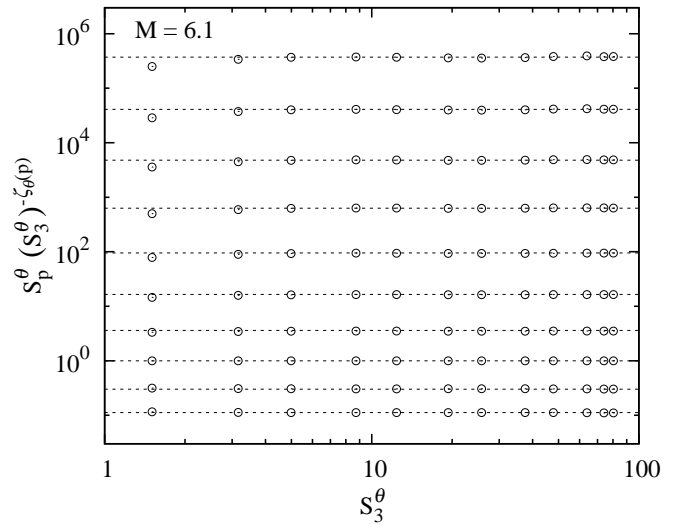} 
\caption{Compensated scalar SFs 
in the $M=6.1$ flow at orders, $p$, from 1 (bottom) to 10 (top).  
Horizontal lines show the quality of the power-law fits. 
For clarity,  SFs at $p=1$ and $2$ are multiplied by 0.3 
and 0.7, respectively.}
\label{fig:scalarm6}
\end{figure}

\begin{figure}
\includegraphics[width=0.5\textwidth]{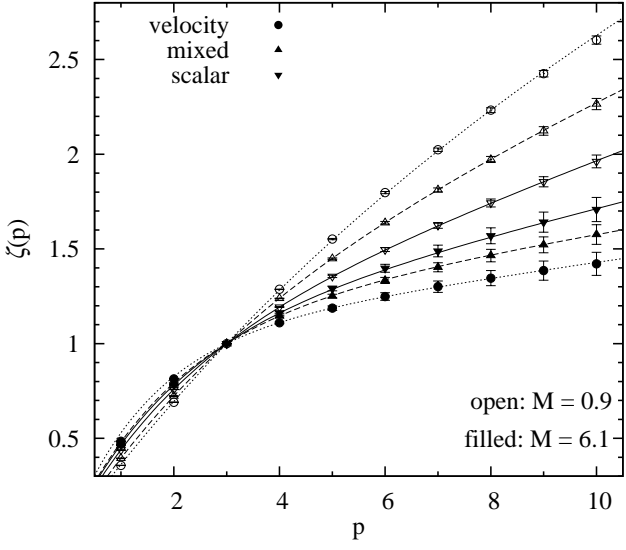} 
\caption{Scaling exponents $\zeta(p)$ 
for SFs at $M=0.9$ and $M=6.1$. 
Note that $\zeta(p) = \xi(p)/\xi(3)$ and 
$\zeta(p) \neq \xi(p)$ when $\xi(3) \neq 1$.  
Lines are predictions of the SL model, 
and parameters used in the 
model are given in Table 1. }
\label{fig:scalings}
\end{figure}

The scaling exponents for the velocity SFs 
decrease steadily and significantly with 
increasing $M$. This is because the 
frequency and the strength of shocks in 
supersonic flows increase with $M$. Shocks 
are strong intermittent structures, which tend 
to decrease the scaling exponents at high 
orders, as illustrated by Burgers turbulence~\cite{fri95}.

In contrast, there is no such effect for the 
scaling exponents of the scalar structures, 
which decrease only slightly as $M$ 
changes from $0.9$ to $6.1$. No 
discontinuous structures like the velocity shocks 
are expected in the scalar field at any $M$. 
As discussed in Ref.~\cite{pan10}, 
velocity shocks can squeeze the 
scalar field, and amplify the 
concentration gradients, leading to edge-like structures 
around the shocks (see Fig.\ 1 in Ref.~\cite{pan10}). 
However, these edges are not 
discontinuities because the 
amplification factor for the 
scalar gradient by a shock is finite, about equal to 
the density jump across the shock. 
The continuity in the scalar field is also expected from the 
conservation of tracer mass. 
The tracer density jump across a shock 
is the same as that of the flow density, 
and thus the concentration, the ratio 
of the two densities, is continuous across the shock. 
The lack of shock-like discontinuities in the 
scalar field is responsible for the different scaling 
behavior of the scalar SFs as a function 
of $M$ from the velocity SFs.

\begin{table}
\begin{center}
\caption{Scaling exponent $\xi (3)$ of 3rd order structure SFs,  
intermittency parameter $\beta$ and the fractal dimension $d$ of MISs}
\label{tbl-1}
\vspace{0.1in}
\begin{tabular}{c|ccc|ccc|ccc}
\colrule
\colrule
$M$&  $\xi_{\rm v} (3) $ & $\beta_{\rm v}$ &  $d_{\rm v}$  & $\xi_{\rm m}(3)$ & $\beta_{\rm m}$ & $d_{\rm m}$&  $\xi_{\theta}(3)$ & $\beta_{\theta}$ & $d_{\theta}$\\ \hline
0.9 &    0.98  &  0.88   &  1.0     & 0.96 &  0.75   &  1.9     &  0.87    &  0.64 &  2.2 \\
1.4 &    1.07  &  0.85   &  1.0     & 0.96 &  0.77   &  1.8     &  0.82    &  0.63 &  2.3 \\
2.1 &    1.18  &  0.77   &  1.2     & 0.96 &  0.78   &  1.5     &  0.78    &  0.65 &  2.2\\
3.0 &    1.22  &  0.63   &  1.7     & 1.01 &  0.72   &  1.7     &  0.82    &  0.62 &  2.2\\
4.6 &    1.22  &  0.54   &  1.8     & 1.03 &  0.64   &  1.8     &  0.91    &  0.62 &  2.1\\
6.1 &    1.22  &  0.52   &  1.8     & 1.07 &  0.61   &  1.8     &  0.96    &  0.61 &  2.0\\
\colrule
\colrule
\end{tabular}
\end{center}
\end{table}
  
We next show that the scaling behaviors of the 
scalar structures in supersonic turbulence can 
be well fit by the SL intermittency model~\cite{she94},  
which has achieved substantial success 
in a wide range of turbulent systems~\cite{mul00, pad04, lev99}.
The model is based on the consideration 
of a hierarchy of structures, characterized 
by ratios of successive SFs, $F_{p}(r) = S_{p+1}(r)/S_{p}(r)$. 
With increasing $p$, $F_{p}(r)$ represents 
structures of higher intensity level, and $F_{\infty}(r)$ 
corresponds to the most intense structures (MISs) 
at the scale $r$.   
The main assumption of the model is the existence 
of a symmetry in the hierarchy,  
\begin{equation}
F_{ p+1}(r) = A_{p} F_{p} (r)^\beta F_{\infty}(r)^{1-\beta},  
\label{eq:symmetry}
\end{equation}  
where $A_p$ is assumed to be independent 
of $r$,  and $\beta$ is referred to as the 
intermittency parameter.

Solving the recursion relation, eq.\ (\ref{eq:symmetry}), 
gives,
 \begin{equation}
\xi (p) = \gamma p + C(1-\beta^p),   
\label{eq:sl}
\end{equation}  
where $\gamma$ is the scaling exponent 
for the MISs, i.e., $F_{\infty} (r) \propto r^{\gamma}$. 
The parameter $C$ has a physical interpretation as 
the codimension of the MISs, defined as $C = 3-d$ 
in 3D with $d$ being the fractal dimension of  
the MISs. 
An appealing feature of the SL model is that, 
through the assumed symmetry, the entire 
hierarchy of structures is related to the MISs. 
Therefore, if applicable, the model provides 
an approach to measure the physical properties 
($\gamma$ and $C$) of the MISs. 
  
If eq.\ (\ref{eq:sl}) applies for $\xi(p)$, 
a similar equation exists for $\zeta(p)$, 
\begin{equation}
\zeta(p) = \gamma' p + C'(1-\beta^p),
\label{eq:zeta}   
\end{equation} 
where $\gamma'= \gamma/ \xi(3)$ and 
$C' = C/ \xi(3)$. Since $\zeta(3) =1$ by 
definition, we have $C' = (1-3 \gamma')/(1-\beta^3)$.  
We will determine $\beta$ and $C'$ from our simulation data. 
Unlike $\gamma$ and $C$, the physical meanings of 
$\gamma'$ and $C'$ are not straightforward when $\xi(3) \ne1$. 
Therefore, after obtaining $\gamma'$ 
and $C'$, we convert them to $\gamma$ 
and $C$ using the measured values 
of $\xi(3)$. In particular, we will calculate 
$d$ by $d = 3- C = 3- \xi(3) C'$. 
We note that the dimension derived 
in Ref.~\cite{pad04} for MIVSs in 
supersonic flows was defined as $3 -C'$.
It is different from our definition of $d$ 
and will be referred to as $d'$.

\begin{figure}
\includegraphics[width=0.5\textwidth]{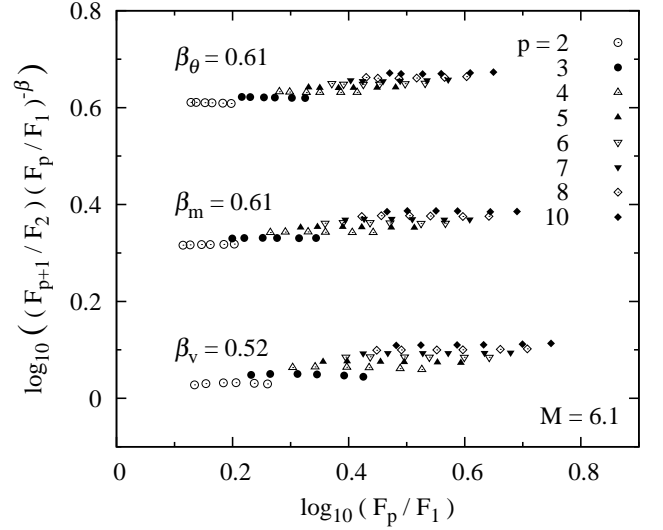} 
\caption{Testing symmetry eq.\ (\ref{eq:symmetry}) in the $M=6.1$ flow.  
For clarity, data points for mixed and scalar structures are 
shifted upward by 0.3 and 0.6, respectively. }
\label{fig:beta}
\end{figure}

Eq.\ (\ref{eq:symmetry}) can be rewritten as~\cite{she01},  
\begin{equation}
\frac {F_{ p+1}(r)}{F_2(r)}  = \frac {A_{p}} {A_1} \left( \frac{F_{p} (r)}{F_1(r)} \right)^\beta.   
\end{equation}  
We can thus examine the validity of the assumed 
symmetry, and hence the applicability of the SL model, 
by checking whether the 
$\log ({F_{ p+1}}/{F_2} )$ -- $\log({F_{ p}}/{F_1} )$ 
curves have the same slope at all orders, $p$~\cite{she01}. 
In particular, if $A_{p}$ is independent of $p$, these curves 
would fall onto the same line. This is indeed the case 
for the velocity structures in incompressible flows~\cite{liu03}. 
The same is found here for the velocity structures at 
$M=0.9$, 1.4 and $2$. In these three cases, we 
estimated $\beta_{\rm v}$ by fitting a single line to 
the data points from all orders, $p$~\cite{liu03}. 

In all other cases, $A_p$ is not constant, and the $\log ({F_{ p+1}}/{F_2} )$ -- $\log({F_{ p}}/{F_1} )$ 
curves at different $p$ are not on the same line. 
However, these curves are generally parallel to 
each other, also confirming the validity of eq.\ (\ref{eq:symmetry}). 
The symmetry test for $M=6.1$ is shown in Fig.\ 3. Similar 
results are found for other $M$. The 6 data points at 
each order correspond to the scale range 
$r \in [12, 64]$. Small variations are observed in 
the slopes at different $p$, and we evaluated 
$\beta$ by averaging the slopes over orders 
up to $p=10$.  

The symmetry test determines the 
parameter $\beta$, and with $\beta$, 
we obtain $C'$ by fitting eq.\ (\ref{eq:zeta}) 
to the measured exponents, $\zeta(p)$. 
The dimension, $d$, is then calculated by 
$d= 3-\xi(3) C'$. The results for $M=0.9$ and $6.1$ 
are shown in Fig.\  2. A good match is 
seen between the model predictions (lines) and 
numerical results (data points). 
The measured values of $\beta$ and $d$ are listed in 
Table 1. Note that, with increasing 
$M$, $\beta_{\rm v}$ decreases significantly, 
while the change in $\beta_{\theta}$ is slight. 
This corresponds to the trend of the scaling exponents 
observed in Fig.\ 2. The parameters obtained in our $M=0.9$ flow agree well 
with those from studies for incompressible turbulence~\cite{liu03, lev99, wat04,rui96}. 
Consistent with Ref.~\cite{pad04}, we find that, with increasing $M$, 
the MIVSs make the transition from filamentary to 
sheet-like, with $d_{\rm v}$ changing from 1 to 
1.8 as $M$ goes from 0.9 to 6.1. 
This range of $d_{\rm v}$ corresponds 
to the change of $d'_{\rm v}$ from 1 to 2, 
in agreement with $d'_{\rm v}$ 
measured in Ref.~\cite{pad04}.

Of particular interest here is the dimension, $d_{\theta}$, 
for scalars. Extensive experimental and numerical 
results have shown that the MISSs in incompressible 
flows are in the form of sheets, known as the cliff 
structures\cite{sre91, pum94}, and the dimension 
$d_{\theta} \simeq 2$ derived in our $M=0.9$ flow is 
consistent with these results. Furthermore, the 
MISSs have a dimension of $\simeq 2$ in 
all our supersonic flows. This result can 
be understood by considering the possible 
effects of compressible modes in shaping the geometry 
of the intense scalar structures. As mentioned earlier, 
strong compressions in supersonic flows 
can produce edge-like scalar structures across shocks~\cite{pan10}. 
The strong shear that usually exists behind shocks
can further make scalar sheets in the narrow 
postshock regions.   
These may provide the main contributions to the MISSs. 
Although formed from a different mechanism, they are 
geometrically similar to the cliff structures found in 
incompressible flows. Therefore whether or not their 
contribution to the MISSs is dominant, $d_{\theta}$ is expected to be 
$\simeq 2$ at all $M$.

The dimension, $d_{\rm m}$, for the mixed structures 
shows a more complicated dependence on $M$ than the
velocity and scalar structures. As $M$ increases from 
$0.9$ and $2$, it decreases from 1.9 to 1.5,  while $d_{\rm v}$ 
increases and $d_{\theta}$ is essentially constant. 
At larger $M$, $d_{\rm m}$ increases again, and 
the MISs for all the three cases have $d$ close to 2 .  

In conclusion, we carried out a systematic 
study for passive scalars  in supersonic 
turbulence.  We find that, with increasing $M$, 
the velocity scalings becomes 
significantly more intermittent, but the degree of 
intermittency in the scalar structures 
increases only slightly with $M$. This is because, 
at any Mach number, the scalar field does 
not have discontinuous structures like velocity shocks. 
The scalar scalings at all Mach numbers are well 
described by the intermittency model of She 
and  L\'{e}v\^{e}que. Fitting the model prediction to the measured 
scaling exponents shows that the most intense 
scalar structures are sheet-like at all Mach numbers.  

\acknowledgements

We acknowledge support from NASA theory grant NNX09AD106.
All simulations were conducted on the ``Saguaro'' cluster operated 
by the Fulton School of Engineering at Arizona State University,
 using the FLASH code, 
a product of the DOE ASC/Alliances-funded Center for 
Astrophysical Thermonuclear Flashes at the University of Chicago.

\end{document}